# From bioinspired multifunctionality to mimumes

*Akhlesh Lakhtakia*
NanoMM—Nanoengineered Metamaterials Group, Department of Engineering Science and Mechanics, The Pennsylvania State University, University Park, PA 16802-6812, USA

**Abstract:** The methodologies of bioinspiration, biomimetics, and bioreplication are inevitably pointing to the incorporation of multifunctionality in engineered materials when designing ever more complex systems. Optimal multifunctionality is also the defining characteristic of metamaterials. As fibrous materials are commonly manufactured from a variety of source materials, mimumes—i.e., microfibrous multifunctional metamaterials—are industrially viable even today, as exemplified by mimumes of parylene C. The microfibrous morphology of mimumes will enhance surface-dominated effects in comparison to those evinced by bulk materials.

## 1. Engineered Biomimicry

The laws of physics hold sway over every biological process just as surely and as completely as over every technological operation. Yet the modification of any biological feature in response to an altered environment leading to the emergence of a new multicellular species is a much lengthier affair than the development of a new product line—especially after the Industrial Revolution, and definitely in the last few decades. In the absence of a prescient selector, natural selection promotes the propagation of the genes of better-adapted mutants, a recognizably new species appearing after some large number of generations. In contrast, feeding on consumer research, technology managers function as prescient selectors to rapidly conceive, design, manufacture, and market improved products. Furthermore, even though the sprouting of a new taxon[1,2] still remains as much of an enigma as that of an entirely novel technoscientific product, the duration of the former is remarkably shorter than of the latter.

Evolution in nature may be much slower than technological evolution, but the former enjoys the advantages of plenitude of mutation. No body of prescient selectors can afford to conduct a large number of experiments in the pursuit of a new or improved product. But numerous mutations arise in nature, albeit over long periods of time. Imparting either insignificant or no reproductive advantage, most mutations do not survive for long. Although proportionally few, the reproductively successful mutations may be regarded as the successful outcomes of a gigantic multidimensional matrix of *experiments*. Accordingly, sapient hominins have long been inspired by attractive outcomes and functionalities evident in plants and animals.

Invisibility from predators provided to animals by skin coloration matched to their environment[3] must have led hominins in the hoary past to camouflage themselves in animal skins while either hunting animals or being hunted by rivals. Today's camouflage uniforms worn by both hunters and soldiers are biomimetic descendants of animal skins.[4] Self-medication by animals to either prevent or treat certain diseases[5,6] must have led medical specialists amongst early humans towards medicinal drugs readily available from natural resources.[7] Bioprospecting continues even more systematically than in the past,[8] the isolation of medicinally effective compounds from the non-human biota being a biomimetic enterprise.

Biomimetics is an engineering methodology whereby a physical and/or chemical mechanism responsible for a specific functionality displayed by a plant or an animal is reproduced for human good. A related engineering methodology is bioinspiration which aims at reproducing a certain outcome of a biological process without reproducing either the underlying mechanism or the biological structure responsible for that outcome. As an example, although aeroplanes were surely inspired by avian flight, aeroplanes do not fly by flapping their wings[9] but birds do. The outcome––i.e., flight—is the same but the mechanisms underlying the flights of a condor and an Airbus A380 are very different. Aeroplanes are bioinspired. In contrast, ornithopters are biomimetic aircraft,[10] as they mimic the flight mechanisms of certain insects.

Both bioinspiration and biomimetics are methodologies of engineered biomimicry.[11] During the last two decades, both have been complemented by a third methodology of engineered biomimicry: bioreplication.[12] This is the direct replication of a structure found in living organisms to thereby reproduce one or more functionalities. As an example, bioreplicated decoys of females of the insect species *Agrilus planipennis* have been shown as highly successful in luring males of the same and related species.[13,14]

Bioreplication research has a Galatean aim. The reader may recall from Greek mythology the story of Pygmalion, a Cypriot sculptor who carved the statue of a woman out of ivory. Falling in love with it, he named it Galatea. As told by the Roman poet Ovid,[15] the goddess Aphrodite granted his desire for a bride who would be the likeness of Galatea, and breathed life into the statue. After coming to life, Galatea married Pygmalion and even bore him a son named Paphos. Manufactured as a replica of a woman, the statue came to exhibit all the functionalities of a female human.

**2. Multifunctionality**

Multifunctionality is commonplace in living organisms.[16-18] Thus, skin contains the organism, defines its shape and size, hosts a variety of sensors, and may be used for camouflage as well as for advertisement. Mouths are used for ingesting food and fluids, releasing sounds, and kissing. The reader can easily find many other examples. Multifunctionality in biology arises from the incorporation of many different materials and many different length scales, whereby surface-dominated and volume-dominated phenomenons are dynamically balanced.[19] This multifunctionality must be distinguished from biochemical multifunctionality, in which large molecules such as proteins and enzymes can participate in diverse chemical reactions by virtue of possessing diverse functional groups of atoms.[20,21]

Multifunctionality engenders economy. When a particular structure can perform two or more distinct functions that are not highly related to each other, fewer structures need to be formed and housed in the organism. Furthermore, fewer structures need to be coordinated by the organism's command center (i.e., the brain). Multifunctionality could even enhance adaptability, in that hitherto-unused (and hithero-obscure) functionalities of a structure could enable the organism to cope with an altered environment.

The economy of multifunctionality is highly attractive to engineers.[22,23] Two canonical ways exist to realize a multifunctional structure. The first canonical way is to fabricate an assembly of monofunctional substructures. A simple example is a Swiss Army knife,[24] to which the human brain has been likened.[25] Two other excellent examples are the combination TV-radio-DVD-CD home entertainment system[26] and the now-ubiquitous smartphone.[27] The monofunctional substructures in such an organizationally multifunctional structure usually share a few modules, which makes them spatially compact.

The second canonical way to engineer a multifunctional structure is to use multifunctional materials.[22,23] Some materials can exhibit multiple functionalities, many of which are useful. An excellent example is lithium niobate.[28] It is polar, i.e., its unit cell has a spontaneous electric dipole moment even in the absence of an applied electric field. When an electric field of sufficient strength is applied, the direction of the spontaneous electric dipole moment is reversed, which is the characteristic property of ferroelectric materials. Heating lithium niobate develops an electric field inside it, as the material is pyroelectric. The material is also piezoelectric, i.e., it can be deformed by the application of an electric field, and the application of mechanical stress creates a dielectric polarization field inside it. The application of mechanical stress also makes lithium niobate optically anisotropic. Thus, this material is photoelastic. The application of a low-frequency electric field linearly changes the optical anisotropy of lithium niobate, which thus exhibits the Pockels effect. The material is optically nonlinear as well. The plethora of useful characteristics and the ability to grow it quite easily lead to the application of lithium niobate in electro-optic modulators, tunable capacitors, piezoelectric sensors, thermal sensors, optical switches, optical waveguides, surface-acoustic-wave devices, and acousto-optic devices.

**3. Metamaterials**

More often than not, a single material is unable to satisfy two or more functional requirements. Recourse then must be taken to composite materials, which are mixtures of two (or more) materials that have quite different properties.[29,30] The mixing can take several forms. In one form, laminas of the constituent materials are made to adhere to each other, giving rise to a laminated composite material. The arrangement of the laminas of different types usually has some order. In another form, granules of one constituent material are dispersed in the other constituent material. The dispersal can be random or possess order over a certain distance. The granules could be identical or dissimilar, and could be randomly oriented or not. When the laminar thickness or the granule size become considerably smaller than some characteristic length scale, the composite material can be considered as effectively a homogeneous material.[29,31] Typically but not always,[32] the effective constitutive parameters (such as Young's modulus and relative permittivity) of the composite material are related to the volume fractions and the constitutive parameters of the constituent materials. Particularly for linear constitutive parameters, additive relations of the kind $P_{eff} = f_1 P_1 + f_2 P_2$ or $1/P_{eff} = f_1/P_1 + f_2/P_2$ based on two capacitors/resistors in parallel/series or series/parallel often emerge—at least in a first analysis.[31] In both relations, $f_1$ and $P_1$ are the volumetric fraction and the constitutive parameter of one constitutive material, $f_2 = 1-f_1$ and $P_2$ are the volumetric fraction and the constitutive parameter of the other constitutive material, and $P_{eff}$ is the effective constitutive parameter of the composite material. For fixed $f_1$, constitutive parameters for two (or more) functionalities would *add* up differently, thereby providing an avenue to simultaneously manipulate two (or more) functionalities for advantage.

Sometimes, the additive relations are violated by a large margin. As an example, suppose that a dielectric composite material is made of a constituent material with relatively high relative permittivity and low frequency-dispersion and another constituent material with relatively low relative permittivity and high frequency-dispersion, in a spectral regime of interest. The effective group speed of electromagnetic waves in the composite can exceed in magnitude the group speed in either constituent material for certain values of $f_1$,[33] which may be useful in reducing information delay in optoelectronic chips.

Sometimes, the composite material displays characteristics that are not shown by its constituent materials. Sculptured thin films[34] provide an example. These composite materials are assemblies of parallel and shaped nanowires grown by evaporating a bulk material, air being the other constitutive material. All nanowires are identical, at least for optical purposes. The nanowire shape imparts optical response characteristics that are not evinced by the bulk material that was evaporated. For instance, a chiral sculptured thin film exhibits the circular Bragg phenomenon[35] arising from the helical shape of its nanowires, which also contributes to a reduction of residual stress in comparison with columnar thin films that are assemblies of straight nanowires. In magnetoelectric laminated composite materials with alternating piezoelectric laminas and magnetostrictive laminas,[36] the interaction of the electric/magnetic field with the strain tensor in two different classes of materials is exploited to deliver magnetoelectric coefficients that can be much higher than those of natural magnetoelectric materials.[37] This is all the more remarkable because neither piezoelectric nor magnetostrictive materials are magnetoelectric.

The two foregoing paragraphs describe a composite material which exhibits[32] "response characteristics that either are not observed in, or are enhanced relative to, the individual responses of its constituent materials, each of which is chemically inert with respect to the others in its immediate proximity." Such a composite material is often called a metamaterial. A clever design of microstructural (or nanostructural) geometry underlies a metamaterial, but multifunctionality does not show up in the working definition quoted.

Nonetheless, multifunctionality appeared in the very first definition of metamaterials given by Walser in 2000.[32,38] Walser later formally defined[39] metamaterials as "macroscopic composites having a manmade, three-dimensional, periodic cellular architecture designed to produce an optimized combination, not available in nature, of *two or more responses* [emphasis in the original] to specific excitation."

So, multifunctionality is intimately tied up with the metamaterial concept. Jettisoning the gratuitous phrase "not available in nature" from Walser's definition (because the veracity of that phrase for any metamaterial can never be conclusively verified), we should define a metamaterial as a composite material that satisfies complex multifunctional requirements by virtue of its morphology being a judicious juxtaposition of sufficiently small cells of different materials, shapes, and sizes.[40]

## 4. Mimumes

Naturally, a variety of bimaterial interfaces exist in multifunctional materials. These materials have to be composites of granules, fibers, and laminas. However, high pressure and/or high temperatures are typically needed to fabricate mechanically robust composite materials exclusively from granules.[41-43] These processing conditions are usually inimical to terrestrial life. Laminated composite materials are also uncommon in the biological word, being found mostly as shells.[44] Biological multifunctionality is commonly exhibited through a fibrous morphology.[45] Fibers provide structural reinforcement[46] as well as pathways for the transport of fluids.[47]

Industries routinely fabricate yarns, textiles, and wools from polymers, rubbers, glasses, metals, and other materials.[48,49] These fibrous materials are used for clothing, personal armor, separation membranes, bioscaffolds, etc.

Fibrosity at the microscale—i.e., microfibrosity—will lead to multifunctionality as follows. The microfibers will provide preferential paths for conduction, diffusion, and other transport mechanisms.[50] Microfibrosity will naturally make surfaces hydrophobic,[51] but suitable surface treatments would impart not only hydrophilicity[52] but also omniphobicity and omniphilicity,[53] if desired. The voids between fibers will provide the microporosity necessary for sensing and microsieving. Controlled microporosity will assist in providing functional gradients[54,55] of thermal, electromagnetic, and elastodynamic impedances. The Bragg phenomenon will be exhibited for electromagnetic[56] and elastodynamic[57] waves, if the gradients are periodic along fixed directions. Microfibrosity will also enhance surface-dominated phenomenons in relation to volume-dominated phenomenons, because the available surface area per unit volume will be higher than in bulk materials.[50] Spatially graded blending of different raw materials during fabrication[55] would also promote multifunctionality.

The conceptualization of **mi**crofibrous **mu**ltifunctional **me**tamaterials—or, mimumes—is thus inspired by biological multifunctionality.[19] The microfibers in a mimume can be made of diverse materials, have diverse lengths and cross sections, and have spatially diverse orientations and distributions. Densely packed microfibers will resemble a bulk material much more than loosely packed microfibers, but in either case the microfibrous morphology will perform differently than any bulk material, because of microporosity.

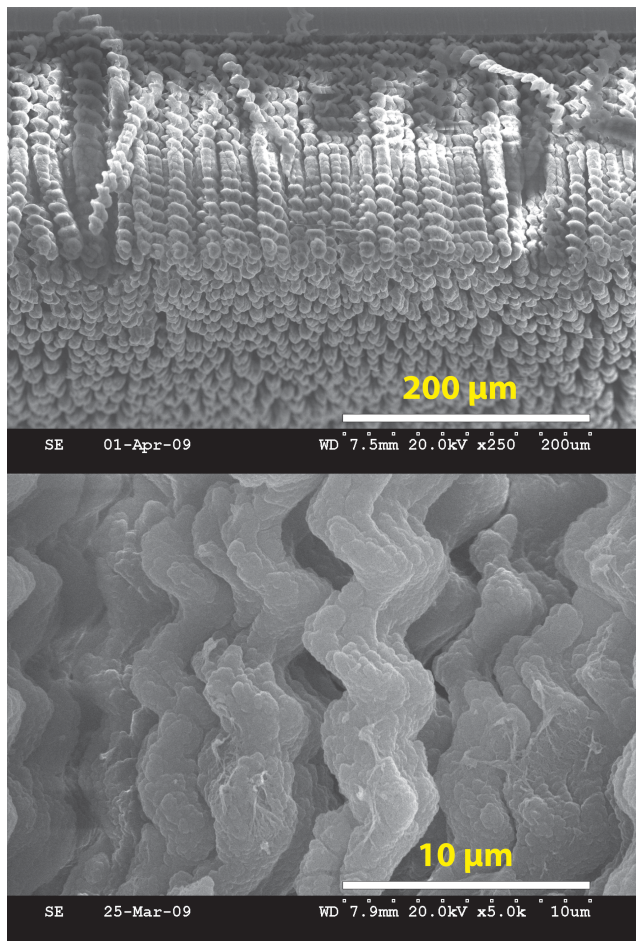

Figure 1. Scanning electron micrographs of two parylene-C mimumes with chiral morphology.

Polymeric mimumes are low-hanging fruit, because textiles made of polymeric yarn are ubiquitous.[49] These textiles are made by electrospinning,[48,58] self assembly,[59] and thermal techniques.[48] Electrospinning is versatile, but fabrication of 3D shapes and control of porosity require further research. Self assembly is inexpensive but does not produce fibers of large diameters and there is little control over fiber diameter and fiber orientation. Thermal techniques have the same disadvantages as self assembly, but their wide industrial use is an advantage.

Scanning electron micrographs of two parylene-C mimumes are shown in Fig. 1. These mimumes were fabricated by a physicochemical thermal evaporation process.[60] Both mimumes are unidirectionally periodic. With periods on the order of 20 μm, the elastodynamic Bragg phenomenon[61] will be displayed at about 35 MHz, and the electromagnetic Bragg phenomenon[35] at about 10 THz. Dyes could be co-deposited,[62] quantum-dot layers could be inserted as phase defects,[63] and liquid crystals could be made to infiltrate[64] these mimumes for optical and electro-optical functionalities. Dye-impregnated mimumes could serve as luminescent concentrators[62] for the harvesting of light by photovoltaic solar cells, while unimpregnated mimumes on solar-cell surfaces could serve as trappers of solar light due to multiple scattering.[65] Parylene-C mimumes have already been shown to act as bioscaffolds for cells,[66] and serum proteins bind to their surfaces.[67] Finally, by using topological substrates, complex, free-standing, flexible, multiscale mimumes have been fabricated.[68] Thus, parylene-C mimumes are strong candidates to exhibit simultaneous ultrasonic, terahertz, light, energy, and biomedical functionalities.[19]

## 5. Concluding Remarks

Endeavors to replicate the features of biological multifunctionality through the methodologies of engineered biomimicry will be natural consequences of research progress. These endeavors are likely to make use of microfibrous morphology for enhanced surface-dominated effects in comparison to those evinced by bulk materials. Thus, the concept of mimumes (microfibrous multifunctional metamaterials) is bioinspired. As fibrous materials are commonly manufactured from a variety of materials for textiles, armor, membranes, and substrates, mimumes are industrially viable even today. This viability is exemplified, either in actuality or potentially, by mimumes of parylene C for ultrasonic, terahertz, light, energy, and biomedical functionalities. The emergence of mimumes will also facilitate the paradigm of concurrent multilevel design of systems for overall performance,[69] which is a rational strategy for designing ever more complex systems.

**Acknowledgment.** The Charles Godfrey Binder Endowment at the Pennsylvania State University is thanked for continued financial support of the author's research.